\documentclass[reprint,twocolumn,amsmath,superscriptaddress,amssymb,aps]{revtex4-1}

\usepackage{graphicx}
\usepackage{graphics}
\usepackage{amsmath}
\usepackage{dcolumn}
\usepackage{bm}

\begin{document}


\title{Hidden chiral symmetry protected $\mathbb{Z}\oplus\mathbb{Z}$ topological insulators in a ladder dimer model}

\author{Jin-Yu Zou}
\author{Bang-Gui Liu}\email{Corresponding author: bgliu@iphy.ac.cn}
\affiliation{Beijing National Laboratory for Condensed Matter Physics, Institute of Physics, Chinese Academy of Sciences, Beijing 100190, China.\\
}%

\date{\today}

\begin{abstract}
We construct a two-leg ladder dimer model by using two orbitals
instead of one in Su-Schrieffer-Heeger (SSH) dimer chain and find
out a chiral symmetry in it. In this model, the otherwise-hidden
additional chiral symmetry allows us to define two chiral massless
fermions and show that there exist interesting topological states
characterized by $\mathbb{Z}\oplus\mathbb{Z}$ and corresponding zero
mode edge states. Our complete phase diagram reveals that there is
an anomalous topologically nontrivial region in addition to the
normal one similar to that of SSH model. The anomalous region
features that the inter-cell hopping constants can be much smaller
than the intra-cell ones, being opposite to the normal region, and
the zero mode edge states do not have well-defined parity each. Finally, we suggest that this ladder dimer
model can be realized in double-well optical lattices, ladder
polymer systems, and adatom double chains on semiconductor surfaces.
\end{abstract}

\pacs{Valid PACS appear here}
\maketitle



The discovery of quantum Hall effect in two-dimensional electron gas
opens a new era in physics\cite{QHE1,QHE2,TKNN}. In such systems,
topology is involved in their electronic phases and phase
transitions. In 2005 Kane {\it et al} \cite{Kane,Kane1} discovered
the time-reversal-invariant topological phase in graphene with
spin-orbit interaction, causing a boom in researching topological
insulator\cite{add1,F and K,LiangFu,Moore,add2,Hasan,Zhang}. Since
topological insulators can be classified according to their
symmetries\cite{BDI}, whether an insulating system is a topological
insulator or not is dependent on the symmetries of its crystalline
and electronic structures. Fortunately, optical lattices engineered
with interfering laser beams allow us to study specific interesting
models which cannot be easily found in nature\cite {double well}.
Such models can possess novel configurations with potentials of
single or multiple periods. Controls of atoms in the $s$ and $p$
orbitals in optical lattices can exhibit many exotic quantum
states\cite {exotic1,exotic2,exotic3,exotic4,exotic5}, including
topological nontrivial states. For instance, there is a topological
insulator phase in an $s$-$p$-orbital ladder reduced from a
two-dimensional double-well optical lattice\cite{optical} and a
topological semimetal in fermionic optical lattice\cite{optical2}.

Simple dimer chain, also known as Su-Schrieffer-Heeger (SSH) model,
is a simple but important model to show typical topological property
of one-dimensional systems\cite{SSH1,SSH2}, although it was
originally proposed to describe one-dimensional polyacetylene\cite
{poly}. The simple dimer chain belongs to BDI class according to its
symmetry\cite {BDI}, and is characterized by the topological
invariant $C\in\mathbb{Z}$.

Here, we construct a two-orbital dimer model, or a ladder dimer
model, which can be realized by optical lattices similar to those
with s and p orbitals\cite{exotic1,optical,optical2}. We find that
the ladder dimer model can possess higher symmetry than the dimer
chain and thus have richer topological phases, beyond BDI class, and
can be characterized by topological invariant
$C\in\mathbb{Z}\oplus\mathbb{Z}$. We plot the complete phase diagram
and work out its zero-mode edge states. Furthermore, we find out the
relationship between this ladder dimer model and massless chiral
fermion. More detailed results will be presented in the following.

\bigskip
{\noindent \bf \large Results}
\bigskip

\textbf{Model and symmetry.} We construct a general two-leg ladder
dimer model by using two orbitals instead of one in the simple SSH
dimer model. Presented in Fig. 1 is a schematic of the ladder dimer
model. Its Hamiltonian can be defined as
\begin{equation}
\label{eq:H_0}
\begin{split}
  H=&\sum_jt_s(a_{jBs}^\dagger a_{jAs}+h.c.)+t_s^{\prime}(a_{jBs}^\dagger a_{j+1,As}+h.c.)\\
  &+t_p(a_{jBp}^\dagger a_{jAp}+h.c.)+t_p^{\prime}(a_{jBp}^\dagger a_{j+1,Ap}+h.c.)\\
  &+t_1(a_{jBp}^\dagger a_{jAs}+h.c.)+t_1^{\prime}(a_{jBp}^\dagger a_{j+1,As}+h.c.)\\
  &+t_2(a_{jBs}^\dagger a_{jAp}+h.c.)+t_2^{\prime}(a_{jBs}^\dagger a_{j+1,Ap}+h.c.)
\end{split}
\end{equation}
where the fermion operators $a_{iPq}$ and $a^\dagger_{iPq}$ ($P$ =
$A$ and $B$, and $q$ = $s$ and $p_x$) are used to define the model. Because we are considering $s$ and $p_x$ orbitals, with the $x$ axis being along the chain, the real-space symmetry of the wave functions implies that the nearest-neighbor hybridization and hopping can be nonzero and the onsite hybridization of the two orbitals needs to be zero. In addition, no next-nearest-neighbor hopping is considered to keep high symmetry, which is consistent with the SSH dimer model.
Defining $\hat{C}_j^{\dagger}$=$[a_{jAs}^{\dagger}
,a_{jBs}^{\dagger} ,a_{jAp}^{\dagger} ,a_{jBp}^{\dagger}]$, we make
a Fourier transformation under periodic boundary condition, and
obtain $\hat{C}_k^{\dagger}=\frac{1}{\sqrt{L}}\sum_jC_j^{\dagger}
e^{ikaj}=[a_{kAs}^{\dagger} ,a_{kBs}^{\dagger} ,a_{kAp}^{\dagger}
,a_{kBp}^{\dagger}]$, where $L$ is the number of the unit cells. As
a result, the Hamiltonian can be simplified into
$H=\sum_k\hat{C}^\dagger_kH_k\hat{C}_k$ with $H_k$ being expressed
as
\begin{equation}\label{H_k}
  H_k=\left(
        \begin{array}{cccc}
          0 & t_s+t_s^{\prime}e^{-ika} & 0 & t_1+t_1^{\prime}e^{-ika} \\
          t_s+t_s^{\prime}e^{ika} & 0 & t_2+t_2^{\prime}e^{ika} & 0 \\
          0 & t_2+t_2^{\prime}e^{-ika} & 0 & t_p+t_p^{\prime}e^{-ika} \\
          t_1+t_1^{\prime}e^{ika} & 0 & t_p+t_p^{\prime}e^{ika} & 0 \\
        \end{array}
      \right)
\end{equation}

\vspace{1cm}
\begin{figure}[!tbp]
\centering
\includegraphics[width=6.5cm]{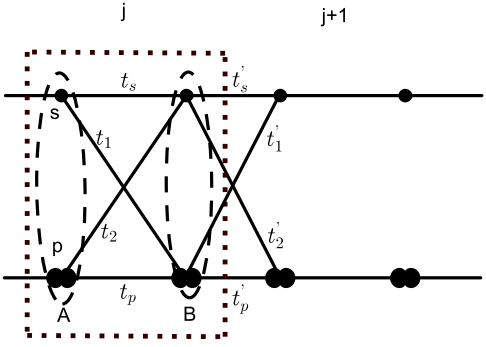}
\caption{A schematic of the ladder dimer model with the hopping
parameters shown.}\label{model}
\end{figure}

Using two sets of Pauli matrices,
$\overrightarrow{\tau}=(\tau_x,\tau_y,\tau_z)$ in the $sp$ orbital
space and $\overrightarrow{\sigma}=(\sigma_x,\sigma_y,\sigma_z)$  in
the AB subspace, we can express the Hamiltonian in a compact form.
Because we are considering real parameters, there should be some
relationships among the parameters\cite{exotic1,optical,optical2}.
If we let the p$_x$ orbitals be in an order like antiferromagnetic
order\cite{exotic1}, we can have $t_1=t_2=\delta$ and
$t_1^\prime=t_2^\prime=\delta^\prime$ for an optical lattice. As a
result, the Hamiltonian can be simplified as
\begin{equation}\label{generalH}
  \begin{split}
  H^a_k=&I \otimes[(t+t^{\prime}\cos{ka})\sigma_x+t^{\prime}\sin{ka}\sigma_y]\\
     &+\tau_z\otimes[(\tau+\tau^{\prime}\cos{ka})\sigma_x+\tau^{\prime}\sin{ka}\sigma_y]\\
     &+\tau_x\otimes[(\delta+\delta^{\prime}\cos{ka})\sigma_x+\delta^{\prime}\sin{ka}\sigma_y]\\
     \end{split}
\end{equation}
where $t=(t_s+t_p)/2$, $t^{\prime}=(t_s^{\prime}+t_p^{\prime})/2$,
$\tau=(t_s-t_p)/2$, and
$\tau^{\prime}=(t_s^{\prime}-t_p^{\prime})/2$. Each of the these
parameters can be nonzero. Consequently, this model has time
reversal symmetry ($T=K$), particle-hole symmetry ($C=I\otimes
\sigma_z K$), chiral symmetry ($S=I\otimes \sigma_z$), and space
inversion symmetry ($R=I\otimes \sigma_x$). It belongs to the BDI
class and is characterized by $\mathbb{Z}$.

\begin{figure}[!tbp]
\centering
\includegraphics[clip, width=6cm]{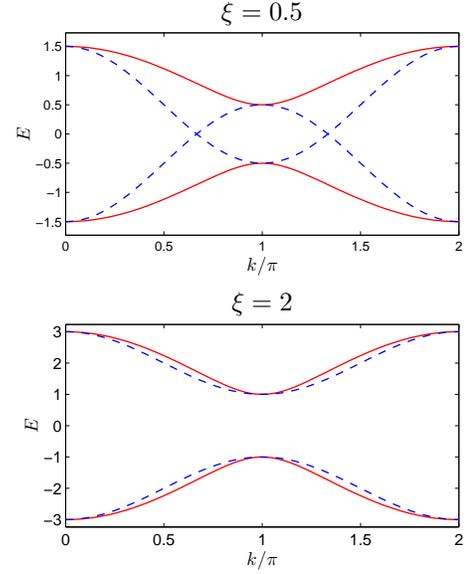}
\caption{The reduced energy spectra of
$E=\pm\sqrt{(\xi+\cos{k})^2+\sin^2{k}}$ with $\xi=0.5$ (upper panel)
and $\xi=2$ (lower panel), in comparison with the curves described
with function $E=\pm(\xi+\cos{k})$.}\label{inverse}
\end{figure}

Introducing the unitary transformation defined by
$U=\exp(-i\alpha\tau_y/2)\otimes I$, we change $H^a_k$ into the
following hamiltonian.
\begin{equation}\label{H_kd}
\begin{split}
  \mathcal{H}^s_k=&U^{\dagger} H^a_k U=(t+t^{\prime}\cos{ka})I\otimes\sigma_x+t^{\prime}\sin{ka}I\otimes\sigma_y\\
  & + [(w+w^{\prime}\cos{ka})\tau_z\otimes\sigma_x+w^{\prime}\sin{ka}\tau_z\otimes\sigma_y]
\end{split}
\end{equation}
where $w=\delta/\sin \alpha$ and $w^\prime=\delta^\prime/\sin
\alpha$, and $\alpha$ (in $[0,\pi]$) is determined by
$\tan\alpha=\delta/\tau=\delta^\prime/\tau^\prime$. Here, the condition $\delta/\tau=\delta^\prime/\tau^\prime$ is required in the diagonalization of the $\tau$ space, which reduces one
freedom in the parameter space, but fortunately, we have only four freedoms in $\mathcal{H}^s_k$, two less than those in $\mathcal{H}^a_k$. This parameter condition can be satisfied by requiring $\delta/\delta^\prime=\tau/\tau^\prime=(t_s-t_p)/(t_s^\prime-t_p^\prime)$ in $\mathcal{H}^a_k$. It is easy to prove that this hamiltonian commutes with $P=\tau_z\otimes I$,
$[\mathcal{H}^s_k,P]=0$. Therefore, we have found a hidden chiral
symmetry in $H^s_k$. It should be pointed out that the hidden chiral symmetry is guaranteed by both the inversion symmetry and the parameter condition. With this additional $P$
chiral symmetry, the hamiltonian can be viewed as massless fermions
and we have four additional symmetries: $T^\prime=PT$,
$C^\prime=PC$, $S^\prime=PS$, and $R^\prime=PR$. Then, we can
introduce the projection operators
\begin{equation}\label{P}
  P_L=\frac{1}{2}(I+\tau_z)\otimes I , P_R=\frac{1}{2}(I-\tau_z)\otimes I
\end{equation}
and write the hamiltonian as a block-diagonal form
\begin{equation}\label{H_kP}
  \mathcal{H}^s_k=\left(
                  \begin{array}{cc}
                    H^s_L(k) & 0 \\
                    0 & H^s_R(k) \\
                  \end{array}
                \right)
\end{equation}
where the $ H^s_L=[(t+ w)+(t^{\prime}+
w^{\prime})\cos{ka}]\sigma_x+(t^{\prime}
+w^{\prime})\sin{ka}\sigma_y$ and
$H^s_R=[(t-w)+(t^{\prime}-w^{\prime})\cos{ka}]\sigma_x+(t^{\prime}-w^{\prime})\sin{ka}\sigma_y$
describe the left and right chiral fermions, respectively. Each of
$H^s_L(k)$ and $H^s_R(k)$ has two independent parameters, and
therefore they are independent of each other.

\textbf{Topological states and  phase diagram.} Diagonalizing the
hamiltonians $H_L$ and $H_R$, we obtain the energy
\begin{equation}
E_\gamma=\pm\sqrt{\Delta_1^2+\Delta_2^2},
 \end{equation}
where $\gamma=(L,R)$, $\Delta_1=(t^{\prime}\pm w^{\prime})\sin{ka}$,
$\Delta_2=(t\pm w)+(t^{\prime}\pm w^{\prime})\cos{ka}$, and the
minus sign corresponds to the left and plus the right. The energy
band structure of the left part becomes gapless if we have
$|\frac{t+w}{t^{\prime}+w^{\prime}}|=1$, and for the right part the
condition is $|\frac{t-w}{t^{\prime}- w^{\prime}}|=1$. Each of the
two parts is a dimer chain. We present in Figure 2 a reduced form of
$E_\gamma$, $E=\pm\sqrt{(\xi+\cos{k})^2+\sin^2{k}}$, with $\xi=\xi_L=(t+
w)/(t^\prime+ w^\prime)$ for the left part and $\xi=\xi_R=(t-
w)/(t^\prime- w^\prime)$ for the right one. It is easy to see the energy structure
inversion in the figure.

\begin{figure}[!tbp]
\centering
\includegraphics[width=7cm]{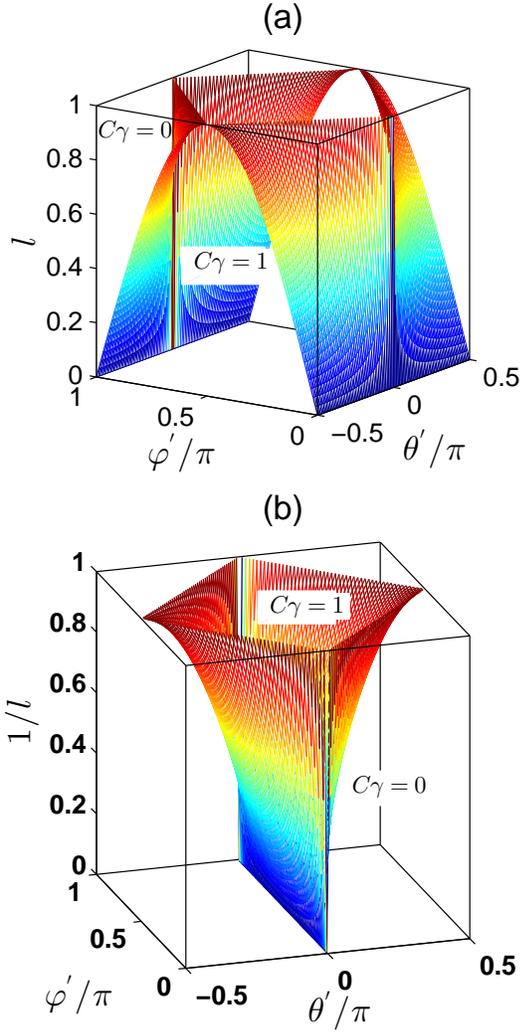}
\caption{The ($l$,$\theta^\prime$,$\varphi^\prime$) phase diagram
for $l|\sin{(\theta^\prime)}|/|\sin{(\varphi^\prime)}|<1$ for
 $l=r/r^{\prime}<1$ (a) and $1/l=r^{\prime}/r<1$ (b). The true ($l$,$\theta$,$\varphi$) diagram is determined by
$l|\sin{(\theta+\pi/4)}/\sin{(\varphi+\pi/4)}|<1$ for the left and
$l|\sin{(\theta-\pi/4)}/\sin{(\varphi-\pi/4)}|<1$ for the right,
with $0\le l<+\infty$. Because of the absolute value sign, the
complete phase diagram can be plotted in terms of
$\theta\in[-\pi/2,\pi/2]$ and $\varphi\in[0,\pi]$.}\label{phase}
\end{figure}

For the hamiltonian $H_\gamma(k)=\vec{h}_\gamma(k)\cdot\vec{\sigma}$
here, the winding number of vector $\vec{h}_\gamma(k)$ is a good
approach to topological characterization because there is no
$\sigma_z$ component\cite{zak,winding},
$\vec{h}_\gamma(k)=\rho_\gamma(k)(\cos{\phi_\gamma(k)},\sin{\phi_\gamma(k)},0)$.
The winding number can be expressed as
\begin{equation}\label{C}
  C_\gamma=\frac{1}{2\pi}\oint\phi^{\prime}_\gamma(k)dk=\frac{\Delta\phi_{\gamma}}{2\pi}
\end{equation}
It should be pointed out that our $C_\gamma$ here is equivalent to
either zero or +1, and the value -1 does not appear because we have
the same coefficients for $\sin ka$ and $\cos ka$ in $H_\gamma(k)$,
which is different from a generalized SSH chain model\cite{add4}.
The total winding number is equivalent to $C=C_L\oplus C_R$. This
definition is well defined because it is gauge invariant. Under this
definition, the left and right parts enter nontrivial topological
states respectively if the following conditions are satisfied.
\begin{equation}\label{condition}
\left\{
\begin{array}{rl}
\displaystyle  |\xi_L|=|\frac{t+w}{t^{\prime}+w^{\prime}}|<1, & {\rm for~ the~ left} \\
\displaystyle  |\xi_R|=|\frac{t-w}{t^{\prime}-w^{\prime}}|<1, & {\rm for~
the~ right}
\end{array}
\right.
\end{equation}
Generally speaking, the two parts become trivial or nontrivial
independently, and therefore the topological invariant belongs to
$\mathbb{Z}\oplus\mathbb{Z}$.

The original four parameters, namely $t$, $w$, $t^\prime$, and
$w^\prime$, are not convenient to describing the topological phase
diagram effectively. Instead, we use $r$, $r^\prime$, $\theta$, and
$\varphi$ in terms of the definition: $t=r\sin\theta$,
$w=r\cos\theta$, $t^{\prime}=r^{\prime}\sin\varphi$, and
$w^{\prime}=r^{\prime}\cos\varphi$. Because
$t^{\prime}=w^{\prime}=0$ means that the model is trivial, we can
assume $r^{\prime} > 0$ and $r\ge0$. The conditions
(\ref{condition}) of the nontrivial topological states are
equivalent to
\begin{equation}
l|\frac{\sin{(\theta+\pi/4)}}{\sin{(\varphi+\pi/4)}}|<1 ~~{\rm or}
~~ l|\frac{\sin{(\theta-\pi/4)}}{\sin{(\varphi-\pi/4)}}|<1.
\end{equation}
where $l=r/r^\prime$. The sign of absolute value guarantees that the
period is $\pi$ for both $\theta$ and $\varphi$. For brevity, we
show a ($l$,$\theta^\prime$,$\varphi^\prime$) phase diagram for
$l|\frac{\sin{(\theta^\prime)}}{\sin{(\varphi^\prime)}}|<1$ in
Figure 3. We have two phases with $C_\gamma=0$ and $C_\gamma=1$ for
$\theta^\prime$. The true phase diagram of the left part can be
obtained by replacing $\theta^\prime$ with $\theta+\pi/4$, and that
of the right part with $\theta-\pi/4$. The whole phase diagram
consists of four phases with $C=(0\oplus 0)$, $(0\oplus 1)$,
$(1\oplus 0)$, and $(1\oplus 1)$. It can be proved that there is no
$C=(0\oplus 0)$ phase for $l<1$ and no $C=(1\oplus 1)$ phase for
$l>1$. It should be pointed out that the $(0\oplus 1)$ phase is topologically
different from the $(1\oplus 0)$ phase. This is because $H^s_L(k)$ is independent of $H^s_R(k)$ in $\mathcal{H}^s_k$. Furthermore, it can be proved that there exist a pair of zero mode edge states at the interface between
the $(1\oplus 0)$ phase of ($\xi_L$,$\xi_R$)=($\xi_1$,$1/\xi_1$) and the $(0\oplus 1)$ phase of ($\xi_L$,$\xi_R$)=($1/\xi_1$,$\xi_1$), where $\xi_1<1$ is assumed. Therefore, the four phases are topologically different from each other.

It is interesting to compare this phase diagram with the
conventional dimer model. Here $r$ describes the hopping parameter
within the unit cell, and $r^{\prime}$ between the unit cells. The
conventional dimer chain is topologically nontrivial when and only
when the hopping within the unit cell is smaller than the hopping
between the unit cells. In contrast, our ladder dimer model can be
topologically nontrivial even when $r$ is larger than $r^{\prime}$.
Furthermore, we find that for $r<r^{\prime}$, the left and right
parts can be topologically nontrivial at the same time, but for
$r>r^{\prime}$, they will never be topologically nontrivial
simultaneously. These happen because for each of the two parts the
hopping terms are determined by two independent parameter freedoms
(not independent parameters).

\textbf{Zero mode edge states.} To explicitly elucidate the
nontrivial topological states, we study the band structures and
explore edge states under open boundary condition. As unveiled by
Delplace {\it et al}\cite{zak}, under the open boundary condition, a
finite single chain with $L_c$ dimers possess two edge states located
at the two ends when $v/v^{\prime}<1-1/(L_c+1)$, where $v$ is
the intra-cell hopping parameter and $v^{\prime}$ the inter-cell
one. For the ladder dimer model, the conditions of edge states are
$l|\frac{t+w}{t^{\prime}+w^{\prime}}|<1-\frac{1}{L_c+1}$ for the left
and $l|\frac{t-w}{t^{\prime}-w^{\prime}}|<1-\frac{1}{L_c+1}$ for the
right. In long chain limit of $L_c$ approaching to $\infty$, these
edge-state conditions are consistent with bulk conditions for
nontrivial topological states. For arbitrary finite $L_c$, we define
$l^\prime=l/[1-1/(L_c+1)]$ and obtain the conditions for the existence
of the zero mode edge states,
\begin{equation}
l^\prime|\frac{\sin{(\theta+\pi/4)}}{\sin{(\varphi+\pi/4)}}|<1
~~{\rm or} ~~
l^\prime|\frac{\sin{(\theta-\pi/4)}}{\sin{(\varphi-\pi/4)}}|<1,
\end{equation}
which takes the same form as the bulk conditions in Eq. (10).
Therefore, the phase diagram of the zero mode edge states can be
obtained by substituting $l^\prime$ for $l$ in Eq. (10) and Figure
3. This makes a clear correspondence between the bulk nontrivial
topological states and the zero mode edge states. It should be
pointed out that there is a special region determined by
$\frac{1}{l^\prime} < |\frac{\sin{(\theta\pm
\pi/4)}}{\sin{(\varphi\pm \pi/4)}}| < \frac{1}{l}$ near the phase
boundary, in which there is no zero mode edge state but there exist
bulk nontrivial topological states. The zero mode edge states will
appear in pair because we have two edges. For
$l^\prime>1$, there can be a pair of zero mode edge states in some
($\theta$,$\varphi$) region and there is not any edge state in the
remaining region. For $l^\prime <1$, there are either one or two
pairs of zero mode edge states for given ($\theta$,$\varphi$)
values. The difference between $l$ and $l^\prime$ can be attributed
to different finite size effect under the open boundary condition
from that under the periodic boundary condition.

\begin{figure}[!tbp]
\centering  
\includegraphics[clip, width=8cm]{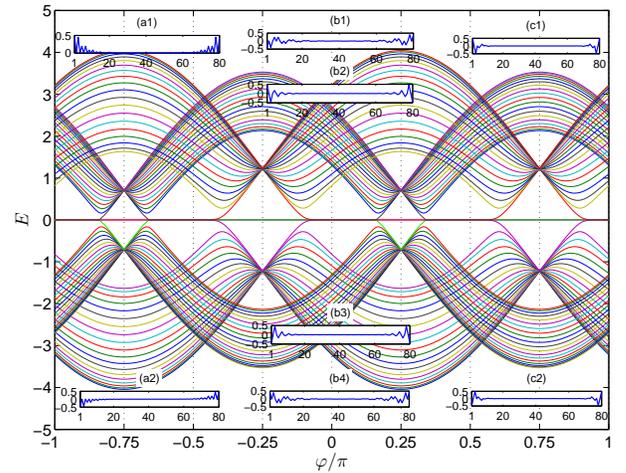}\\
\caption{The energy levels as functions of the parameter $\varphi$
with $L_c=20$, $r=1$, $r^{\prime}=2$, and $\theta=5\pi/12$. At least
one of the two parts is topologically nontrivial because of $l=1/2$
and $l^\prime=21/40$. The inserts (a1) and (a2) are the two zero
mode edge states for $\varphi=\pi/4$ ($-3\pi/4$ also), corresponding
to the nontrivial right part. The inserts (c1) and (c2) are the two
edge states for $\varphi=-\pi/4$ ($3\pi/4$ also), corresponding to
the nontrivial left part. The (b1-b4) shows the four edge states for
$\varphi=0$ ($\pm \pi/2$ and $\pm \pi$ also), which corresponds to
the situation in which both of the two parts are topologically
nontrivial. The edge states have well-defined parity and appear in
pair.}\label{edge}
\end{figure}

\begin{figure}[!tbp]
  \centering
  \includegraphics[clip,width=8cm]{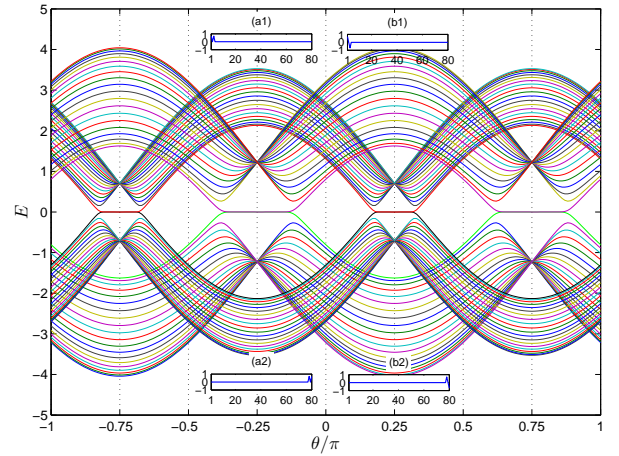}
\caption{The energy levels as functions of the parameter $\theta$
with $L_c=20$, $r=2$, $r^{\prime}=1$, and $\varphi=5\pi/12$. In this
case of $l=2$ and $l^\prime=21/10$, at most one of the two parts can
be nontrivial. The inserts (a1) and (a2) are the two zero mode edge
states for $\theta=-\pi/4$ ($3\pi/4$ also), corresponding to the
nontrivial right part. The inserts (b1) and (b2) are the two edge
states for $\varphi=\pi/4$ ($-3\pi/4$ also), corresponding to the
nontrivial left part. The edge states appear in pair and have no
defined parity.}\label{energy1}
\end{figure}

In order to show the edge states more clearly, we present in Figure
4 the energy levels as functions of the parameter $\varphi$ with
given parameters of $L_c=20$, $r=1$, $r^{\prime}=2$, and
$\theta=\frac{5}{12}\pi$. Here, because of $l=1/2$ and
$l^\prime=21/40$, it can be inferred from Eqs. (10) and (11) that at
least one of the two parts are topologically nontrivial, and there
are either 2 or 1 pairs of zero mode edge states. For the $\varphi$
values near both $\varphi=+\pi/4$ and $-\pi/4$, there are one pair
of zero mode edge states with opposite parity. In contrast, there
are two pairs of zero mode edge states near $\varphi=0$, with one
pair of them belonging to the left part and the others to the right
part. Each of these edge states has well-defined parity. The
transition points are determined by $|\sin(\varphi\pm
\pi/4)|>\sqrt{3}l^\prime /2=21\sqrt{3}/80$.

Presented in Figure 5 are the energy levels as functions of the
parameter $\theta$ with $L_c=20$, $r=2$, $r^{\prime}=1$, and
$\varphi=\frac{5}{12}\pi$. In this case, we have $l=2$ and
$l^\prime=21/10$, and there are at most one pair of zero mode edge
states according to Eq. (12). It is clear in Figure 5 that there is
no edge state near $\theta=0$ and equivalent points, and there are
one pair of zero mode edge states near $\theta=\mp \pi/4$ and
equivalent points. Because being degenerate in energy, they have no well-defined parity, in contrast to
those in the case of $l^\prime<1$. The transition points are determined by
$|\sin(\theta\pm \pi/4)|<1/(2l^\prime)=5/21$. The two edge states can be transformed into each other by the space
inversion.

\bigskip
{\noindent \bf \large Discussion}
\bigskip

\textbf{Zak phases.} In addition to the winding number, the Zak phase
can be used to characterize the ladder dimer model. The Zak phase for $\gamma$ ($L$,$R$)
can be defined as
\begin{equation} \Phi^\gamma_Z=i\oint dp\langle
u^\gamma_q|\partial_qu^\gamma_q\rangle,
\end{equation}
where the $|u^\gamma_q\rangle$ are the Bloch wave functions, with
$\gamma$ being both $L$ and $R$. For such one-dimensional models, $\Phi^\gamma_Z$ is
defined up to $2\pi$ due to the gauge transformation. For each of $L$ and $R$, the Zak phase is quantized so that
$\Phi^\gamma_Z=0$, $\pi$ Mod[$2\pi$], because of the
inversion symmetry. Such Zak phase has been
measured in the case of simple SSH model\cite{add3}. Defining
$Z_\gamma=\Phi^\gamma_Z/\pi$, we have $Z_\gamma=0$ and 1 accordingly.
It can be proved in terms of definitions (8) and (12) and the gauge invariance that $Z_\gamma$ is equivalent to $C_\gamma$. As a result, we can also use the Zak phases to characterize the topological phases of the model.

\textbf{High chiral symmetry and $\mathbb{Z}\oplus\mathbb{Z}$ uniqueness.} At first sight there would be many high-symmetry ladder dimer models from the SSH dimer mode, but actually it is not the case. First of all, both the on-site hybridization and the next-nearest-neighbor hopping needs to be forbidden to keep the $T$, $C$, $S$, and $R$ symmetries and the hidden chiral symmetry $P$. This requires that any vertical rung is absent in Fig. 1, in contrast with multi-leg extension with vertical rungs\cite{zak}. In addition, there is a restrictive parameter condition between the inter-orbital and the intra-orbital hopping parameters. The hidden chiral symmetry is necessary to the additional $T^\prime$, $C^\prime$, $S^\prime$, and $R^\prime$ symmetries and the resultant $\mathbb{Z}\oplus\mathbb{Z}$ topological insulators in the ladder dimer model. The two conditions for nontrivial topological phases in the ladder dimer model, $\xi_L<1$ for the left part and $\xi_R<1$ for the right in (9), are each similar to that in the SSH dimer model. The unitary transformation $U$ allows one more parameter freedom in the original hamiltonian $H^a_k$. Actually, one cannot achieve higher chiral symmetry in such two-leg ladder dimer model from the SSH model. Therefore, in this sense, our ladder dimer model possessing the high chiral symmetry (including the hidden chiral symmetry) and the $\mathbb{Z}\oplus\mathbb{Z}$ topological insulator phases is unique up to a unitary transformation.

\textbf{Experimental realization.} Now we
address how to realize the ladder dimer model in (1). Inspired by
earlier work\cite{exotic1,optical}, we propose to realize the
one-dimensional model with a two-orbital optical lattice with the
optical potential given by
\begin{equation}\label{potential}
\begin{split}
  V(x,y)&=V_{x1}\sin^2{kx}+V_{x2}\sin^2{(2kx+\frac{\pi}{2})}\\
  &+V_{y1}\sin^2{ky}+V_{y2}\sin^2{(2ky+\frac{\phi}{2})},
\end{split}
\end{equation}
where we keep $V_{y1,2}\gg V_{x1,2}$. This optical lattice has a
double-well structure in the $y$ direction and causes the particles
to transit alternately between the sub-wells with two different
tunneling barriers ($B_1$ and $B_2$) in the $x$ direction. The
tunneling barrier between the two wells in the $y$ direction is much
larger than $B_1$ and $B_2$ in the $x$ direction. Consequently, the
low-energy physics of the two-dimensional system (13) reduces to a
two-leg ladder model with alternate barriers between sub-wells in
the $x$ direction. We can control parameters $\phi$ and $V_{y1,2}$
to tune the well depth of the two legs, and change $V_{x1,2}$ to
tune the depth of the sub-wells in each leg. Letting the $s$ orbital
of the upper leg have the same energy as the $p_x$ orbital of the
lower leg, we realize the ladder dimer model showed in Fig. 1 and
defined in Eq. (1) by filling the low-lying levels of the relevant
$s$ and $p_x$ orbitals with a single species of
fermions\cite{exotic1,optical}. Making the p$_x$ orbitals be in an
order like antiferromagnetic order\cite{exotic1}, we can obtain
$t_1=t_2=\delta$ and $t_1^\prime=t_2^\prime=\delta^\prime$ and
simplify the Hamiltonian (1) into Eq. (3). In addition, this model
should be realized in ladder polymer systems or adatom double chains
on appropriate semiconductor surfaces.

\textbf{Conclusion.} In summary, we have constructed a ladder dimer
model with higher symmetry, defined in Eq. (3), than the usual SSH
dimer chain model. In this case, there is a hidden chiral symmetry
between the two orbitals that allows us to define two chiral
massless fermions. Furthermore, we show that the ladder dimer model
can exhibit interesting topological states characterized by
$\mathbb{Z}\oplus\mathbb{Z}$ and have zero mode edge states,
assuming a one-dimensional $\mathbb{Z}\oplus\mathbb{Z}$ topological
insulator. We find out complete phase diagram for the bulk
topological states and zero mode edge states. Importantly, the phase
diagram reveals that there exist nontrivial topological states in
not only the normal region with $r^\prime \ge r$ but also the
anomalous region with $r^\prime <r$, where $r^\prime$ is the
inter-cell hopping constant and $r$ the intra-cell one, in contrast
with the usual dimer chain model which shows nontrivial topological
states only in the $r^\prime \ge r$ region. We also find that the
zero mode edge states have well-defined parity
in the normal region, but do not in the anomalous region. Finally,
we suggest that this ladder dimer model can be realized in
double-well optical lattices, ladder polymer systems, and adatom
double chains on semiconductor surfaces.

%
%

\vspace{0.5cm}
\noindent \textbf{\large Acknowledgments}\\
%
This work is supported by Nature Science Foundation of China (Grant
No. 11174359), by Chinese Department of Science and Technology
(Grant No. 2012CB932302), and by the Strategic Priority Research
Program of the Chinese Academy of Sciences (Grant No. XDB07000000).

\vspace{0.5cm}
\noindent \textbf{\large Author contributions}\\
BGL supervises the project. JYZ did all the derivation. Both carried
out the analysis and wrote the manuscript.

\vspace{0.5cm}
\noindent \textbf{\large Additional information}\\
Competing financial interests: The authors declare no competing
financial interests.

\end{document}